\documentclass[ reprint, superscriptaddress,showpacs,preprintnumbers,
 amsmath,amssymb, aps, pra, longbibliography]{revtex4-1}
\usepackage{color}
\usepackage{ulem}
\usepackage{mathtools}
\usepackage{graphicx}
\usepackage{subfigure}
\usepackage{multirow}
\renewcommand{\thesubfigure}{\alph{subfigure}}
\makeatletter
\renewcommand{\@thesubfigure}{(\thesubfigure)\space}
\makeatother
\usepackage{dcolumn}
\usepackage{bm}
\usepackage[colorlinks, linkcolor=blue, citecolor=blue, urlcolor=blue]{hyperref}
\usepackage{geometry}
\begin{document}
\title{Entanglements and correlations of one-dimensional quantum spin-1/2 chain with anisotropic power-law long range interactions}
\author{Jie Ren}
\email{jren@cslg.edu.cn}
\affiliation{Department of Physics, Changshu Institute of Technology, Changshu 215500, China}

\author{Wen-Long You}
\affiliation{College of Science, Nanjing University of Aeronautics and Astronautics, Nanjing, 211106, China}
\affiliation{School of Physical Science and Technology, Soochow University, Suzhou, Jiangsu 215006, China}

\author{Xiaoqun Wang}
\email{xiaoqunwang@sjtu.edu.cn}
\affiliation{Key Laboratory of Artificial Structures and Quantum Control of MOE,
Shenyang National Laboratory for Materials Science, School of Physics and Astronomy,\\
Tsung-Dao Lee Institute, Shanghai Jiao Tong University, Shanghai 200240, China}
\affiliation{Collaborative Innovation Center for Advanced Microstructures, Nanjing University, Nanjing 210093, China}
\affiliation{Beijing Computational Science Research Center, Beijing 100084, China}

\date{\today}
\begin{abstract}
The correlations, entanglement entropy, and fidelity susceptibility are calculated for a one-dimensional spin-1/2 XXZ chain with anisotropic power-law long range interactions by employing the density matrix renormalization group method. In particular, this long-range interaction is assigned to ferromagnetic for transversal components, while it can be either ferro- or antiferromagnetic for the longitudinal spin component. Two ground-state phase diagrams are established versus the anisotropy of the interactions which not only changes the phase boundaries of the counterparts with short-range interactions, but also leads to the emergence of exotic phases. We found that the long-range interactions of the $z$-component results in a Wigner crystal phase, whereas the transversal one may break a continuous symmetry, resulting in a continuous symmetry breaking  phase.
\end{abstract}
\maketitle
\section{Introduction}
The quantum phase transition (QPT) and quantum critical phenomena are generally important in understanding novel properties involved in strongly correlated systems, such as quantum magnetic materials. Usually, short-range interactions, e.g., nearest neighbor and next nearest neighbor interactions, are considered to be sufficient for appropriate descriptions on the major magnetic properties of those systems~\cite{Sachdev,XWang2000,Luo2017,You19,WN19,Luo2019}.
However, there actually exist several types of long range interactions such as the Coulomb interaction $1/r$~\cite{Saffman}, the dipole-dipole interaction $1/r^3$~\cite{Lahaye,Deng,Yan}, and the van der Waals interaction $1/r^6$~\cite{Saffman} in some complicated compounds, where relevant electrons are in higher orbits of atoms with lower symmetries subject to crystal field effects.
Moreover, in recent years, some long-range interactions have been generated in ultracold atomic systems with the optical lattices or trapped ions. For instance, a power-law Ising interaction $1/r^\alpha$ with an adjustable exponent $0 <\alpha< 3$ has been realized in trapped ions~\cite{Britton,Islam,Gorshkov,Jurcevic}.
This kind of experimental progress has greatly stimulated theoretical studies on possible novel effects particularly resulting from long-range interactions ~\cite{W,Koffel,Sun01,Zhu,gong16,gong17,gong17L,gong16R,Frerot,Vanderstraeten}.
In particular, a transition was revealed by the calculation of the entanglement for a long-range ($\sim r^{-\alpha}$) antiferromagnetic Ising chain~\cite{Koffel},  and is affirmed further by the fidelity susceptibility, being second-order for all $\alpha$~\cite{Sun01,Zhu}.
Moreover, by combining the linear spin-wave theory, field theory approach and density-matrix renormalization-group (DMRG)~\cite{white,KWHP,U01,U02,McCulloch}, effects of the long range interactions on local correlation functions, entanglement entropy and central charge are investigated for both spin-1/2~\cite{gong17} and spin-1~\cite{gong16} to 
await experimental observation. In addition, one also finds that long-range interactions and long-range hopping may lead to drastic effects on the many-body localization in a one-dimensional (1D) spinless fermion system~\cite{Nag2019}, which essentially corresponds to a $XY$ type of long range spin interaction. In this regard, the anisotropic long-range spin interaction can be anticipated to give rise to more effects on quantum transitions.

In this paper, we study a ID spin-1/2 XXZ system with anisotropic power-law long range interactions in terms of the entanglement entropy, fidelity susceptibility, and  correlation functions by performing DMRG calculations. Phase diagrams are established with respect to the power exponents and the anisotropy of interactions. In the following, Sec \ref{sec:Hamiltonian} presents the Hamiltonian in our studies. The details on DMRG calculations and the definitions of those calculated quantities are discussed in Sec. \ref{sec:Measurements}. Numerical results are shown in Sec \ref{sec:Results} with further discussions given in the last section.

\section{Hamiltonian}
\label{sec:Hamiltonian}
In the paper, we consider the following spin-$1/2$ chain with anisotropic long-range interactions, and its Hamiltonian is given by:
\begin{eqnarray}
\label{Hamiltonian}
H=\sum_{j>i}\{\frac{J_{xy}}{|i-j|^\alpha}(S^x_iS^x_{j}+S^y_iS^y_{j})+\frac{J_z}{|i-j|^\beta}S^z_iS^z_{j}\},
\end{eqnarray}
where $i$ and $j$ are the sites of one dimensional lattice, and $S^{\gamma}=\sigma^{\gamma}/2$ with $\gamma =x,y$, or $z$, setting $\hbar=1$  and $\sigma^\gamma$ being the Pauli matrices. Interactions between two spins separated  by a distance of $r=|i-j|$ decay as $r^{-\alpha}$ for both $x$ and $y$ components of spins, but as $r^{-\beta}$ for the $z$ direction. As usual, the parameters $\alpha,\beta$ are both taken positive, while $J_{xy}=-1$ is set up for the simplicity so that $J_z$ readily stands for an anisotropic parameter involved in the establishment of the phase diagram.

For this system, in the limit of $\alpha,\beta\rightarrow+\infty$, the Hamiltonian is reduced to describe a spin-1/2 anisotropic chain with the nearest-neighbor interaction. It turns out that the system involves a ferromagnetic (FM) phase for $J_z< -1$, whereas a gapful
antiferromagnetic (AFM) phase can be shown for $J_z> 1$. Furthermore, in the region of $-1<J_z\leq1$, the system displays an $XY$ phase where quantum fluctuations exclude the existence of any long-range order but correlation functions behave as a power-law decay of the distance characterized as in the Luttinger liquid.

For more general values of  $\alpha$ and $\beta$, long range interactions may result in different features for those phases, which are expected also to be properly characterized by long-distance correlation functions as exploited below.

\section{Measurements and Method}
\label{sec:Measurements}

Thanks to the DMRG method\cite{white,KWHP,U01}, the ground state properties of quasi-one-dimensional systems can be calculated with
very high accuracy. For the present studies of Hamiltonian (\ref{Hamiltonian}), we adopt both infinite-size DMRG (iDMRG)~\cite{McCulloch} and finite-size DMRG, which are based on matrix product states~\cite{U02}. The number of eigenstates for the reduced matrix is kept up to $m=400$ in the truncation of bases, which allows the truncation error to be smaller than $10^{-9}$. In our calculations where finite-size DMRG algorithm, we handle the long range interaction with directly using as a summation over matrix product of operators (MPOs) rather than the summation of finite exponential terms with MPOs~\cite{Vidal}, which inevitably introduces additional systematic error otherwise.
Our codes are mainly based on iTensor C++ library~\cite{tesnor}.

 Since the $z$-component of the total spins for the present system commutes with the Hamiltonian (\ref{Hamiltonian}), the ground-state energy is obtained by comparing the lowest energies for each subspace of $S^z_t=\sum_{i=1}^L \langle S^z_i\rangle$. We found that the ground state resides in the sector of either $S^z_t=0$ or $S^z_t=L/2$. To examine the reliability of our numerics, we also perform the finite-size DMRG with varying the number of states in the truncated bases. Once the ground state energy and the corresponding ground state are identified accurately, the first excited state and the corresponding energy (gap) can be determined similarly as orthonormalized to the ground state.

For a quantum many-body system, the entanglement entropy (EE) can be extracted from the ground state wavefunction $|\psi_0\rangle$ properly to characterize the quantum phase transition induced by the interaction or external fields. Usually, one may separate a given Hamiltonian into two subsystems $A$ and $B$, and compute the reduced density matrix for part $A$ by partially tracing over the degree of freedom of the subsystem $B$, which can be written formally as
$$\rho_{A}=\textrm{Tr}_{B}(|\psi_0\rangle \langle\psi_0|).$$
Then, the entanglement entropy measuring the entanglement between parts $A$ and $B$ is given by
\begin{eqnarray}
\label{eq2} S_A=-\textrm{Tr}(\rho_{A}\ln\rho_{A}).
\end{eqnarray}
which is evaluated in terms of the eigenvalues of $\rho_{A}$ feasibly in DMRG calculations.
For a one-dimensional short-range interacting system with an open boundary condition (OBC), the conformal field theory (CFT) suggests that the entanglement entropy for the subsystem $A$ with size $l$ possesses the following finite-size $L$ scaling behavior~\cite{Cardy}
\begin{eqnarray}
\label{eqSl} S_l=\frac{c}{6}\ln [\frac{L}{\pi} \sin(\frac{\pi l}{L})]+ S_0,
\end{eqnarray}
where $c$ is the central charge which usually has different values for different phases and $S_0$ is a non-universal constant. This scaling behavior has been employed to explore the critical entanglement of defects \cite{Zhao2006} and Gaussian transition\cite{Hu2011}. In this paper, we will show that this scaling behavior is applicable to a case associated with long-range interactions.

\section{Results}
\label{sec:Results}
\subsection{$1/\alpha =0$}
Now we first consider the case of $\alpha=\infty$, which implies that only the nearest-neighbor term of $xy-$interaction survives. It turn out that the long-range interaction for $z-$component governed by $\beta$ may result in novel properties in competition with the $xy-$components.
In this case, Hamiltonian (\ref{Hamiltonian}) can be recast to describe a one-dimensional interacting spinless fermionic chain via the Jordan-Wigner transformation:
\begin{eqnarray}
\label{JWt}
S^z_i&=&\frac{1}{2}-c_i^\dagger c_i,\nonumber\\
S^+_i&=&e^{i\pi \sum_{j=1}^{i-1}c_i^\dagger c_i} c_i,\nonumber\\
S^-_i&=&e^{i\pi \sum_{j=1}^{i-1}c_i^\dagger c_i} c_i^\dagger,\nonumber
\end{eqnarray}
where $S^{\pm}_i$=$S^x_i$ $\pm$ $i S^y_i$ are the raising and lowering spin operators. Subsequently, the ferromagnetic $J_{xy}-$term thus simply represents the hopping of fermions, while the $J_{z}-$term stands for the density-density interactions of fermions, which can be either attractive for $J_z<0$ or repulsive for $J_z>0$. One may expect that this density-density interaction results in quantum transitions for different $\alpha$ and $\beta$.

\begin{figure}[t]
\includegraphics[width=0.48\textwidth]{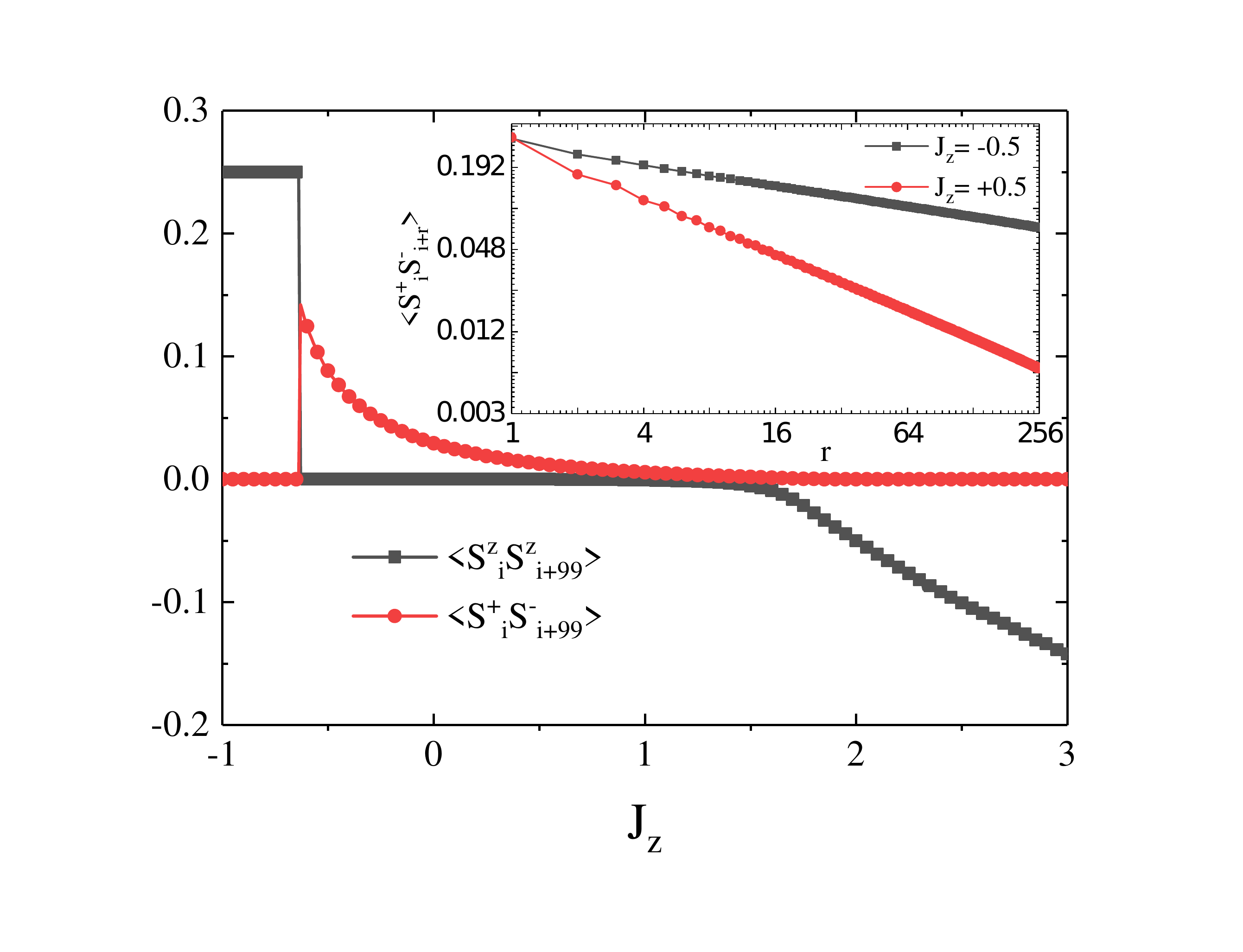}
\caption{\label{fig1} (Color online) Correlation functions $\langle S^+_{i}S^-_{i+r}\rangle$ and $\langle S^z_{i}S^z_{i+r}\rangle $ are plotted as a function of $z$-component interaction $J_{z}$ for $\alpha=\infty$, $\beta=2$ and $r=99$. Inset: a log-log plot for $\langle S^+_{i}S^-_{i+r}\rangle $ as a function of $r$ when $J_z=\pm 0.5$.}
\end{figure}

To explore this, we compute the correlation functions between two spins at $i$ and $j$ with a distance of $r=|i-j| $ and for $\beta$ = 2 with using the iDMRG algorithm. Figure \ref{fig1} shows results for $r=99$.  One can see that when $J_z<-0.636$, the transverse correlation $\langle S^+_{i}S^-_{i+99}\rangle=0$ and the longitudinal correlation $\langle S^z_{i}S^z_{i+99}\rangle =1/4$, implying that the system is in the FM phase, and then $\langle S^+_{i}S^-_{i+99}\rangle$ suddenly jumps to a positive value at $J_z=-0.636$ and $\langle S^z_{i}S^z_{i+99}\rangle$ drops to zero simultaneously. This discontinuity indicates that the ground state undergoes a first order transition from the FM phase into the $XY$ phase. This discontinuous feature is thus utilized here to determine the critical values of $\beta$ and $J_z$ for the quantum phase transition between the $XY$ and FM phases.

\begin{figure}[h]
\includegraphics[width=0.49\textwidth]{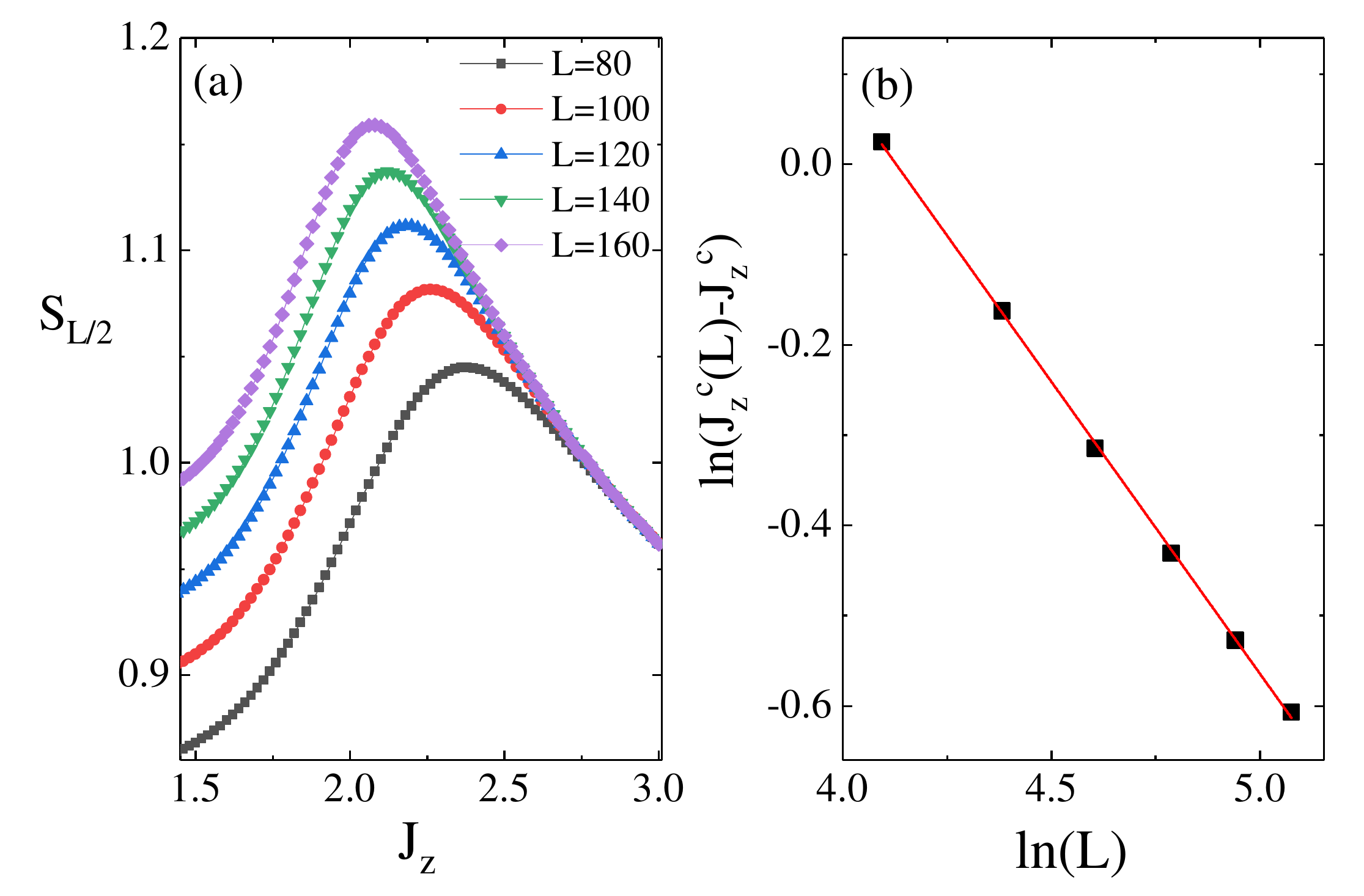}
\caption{\label{fig2} (Color online) (a) Entanglement entropies are plotted as a function of $z$-component interaction $J_{z}$ for various system sizes with $\alpha=\infty$ and $\beta=2$. (b) The peak positions of $S_{L/2}$ versus system sizes $L$.
}
\end{figure}
Moreover, as $J_z$ further increases, the transverse correlation $\langle S^+_{i}S^-_{i+99}\rangle$ gradually reduce to zero, while the longitudinal correlation $\langle S^z_{i}S^z_{i+99}\rangle$ turns to negative for $J_z\gtrsim 3/2$, which signals that the system is driven into a AFM phase. A little scrutiny reveals that the transverse correlation $\langle S^+_{i}S^-_{i+r}\rangle$ satisfies a power-law decay with the distance $r$~\cite{gong17}, as manifested in the inset of Fig. \ref{fig1}.
To determine the critical point at the transition between the $XY$ phase and AFM phase more precisely, we also calculate the von Neumann entropy, i.e. entanglement entropy, for the right part apart from the rest for the chain with using the finite-size DMRG algorithm. The entanglement entropy is shown in Fig. \ref{fig2} as a function of  $J_z$ with $\beta=2$ for different sizes of the chain. With increasing $J_z$, the EE increases first and then declines. The peak becomes more pronounced for a larger size $L$ and the location of the peak moves to a lower value of $J_z$, characterizing a transition between the $XY$ phase and the AFM phase~\cite{Wang}. According to the finite-size scaling theory~\cite{Fisher,Barber83}, it is expected that the position of the pseudo-critical point for a finite-size system approaches the true critical point as $L$ $\to$ $\infty$. For relevant operators in the driving Hamiltonian on sufficiently large-size systems, i.e.,
$\nu$$d$$<$2, where $\nu$ is the critical exponent of the correlation length and $d$ the dimensionality of the system, the leading term in the expansion of pseudo-critical point obeys
\begin{eqnarray}
\label{eq3} |J_z^c(L)-J_z^c(\infty)|\propto L^{-1/\nu},
\end{eqnarray}
where $J_z^c (\infty)$ is the critical value for the thermodynamic limit. Such algebraic convergence can be accelerated considerably by some elaborated strategies~\cite{Roncaglia}. We obtain that $J_z^{c}=1.520$ and $\nu=1.695$ for the present case consistent with the inflection point of the correlations shown in Fig. \ref{fig1}. We note that the scaling behavior of Eq. (\ref{eq3}) with $L$ is also valid for the maximum of fidelity susceptibility defined in Eq.(\ref{eq5})~\cite{You2011} (see below).

\begin{figure}[b]
\includegraphics[width=0.5\textwidth]{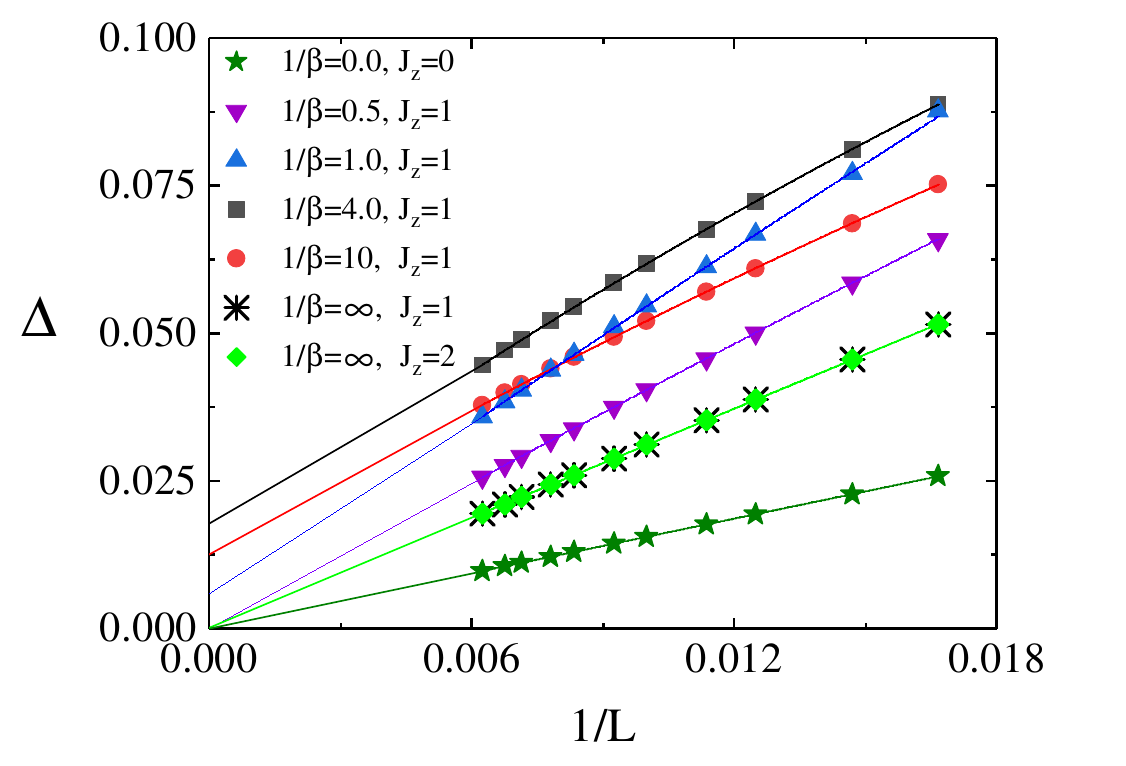}
\caption{\label{fig3} (Color online) Finite size scaling of the energy gap $\Delta$ with various $\beta$ and $J_z$. Symbols show numerical results obtained by DMRG calculations and solid lines are fits of the data by quadratic polynomials in $1/L$. The results for $J_z=0$ is also plotted as for comparison.}
\end{figure}

\begin{figure}[h]
\includegraphics[width=0.48\textwidth]{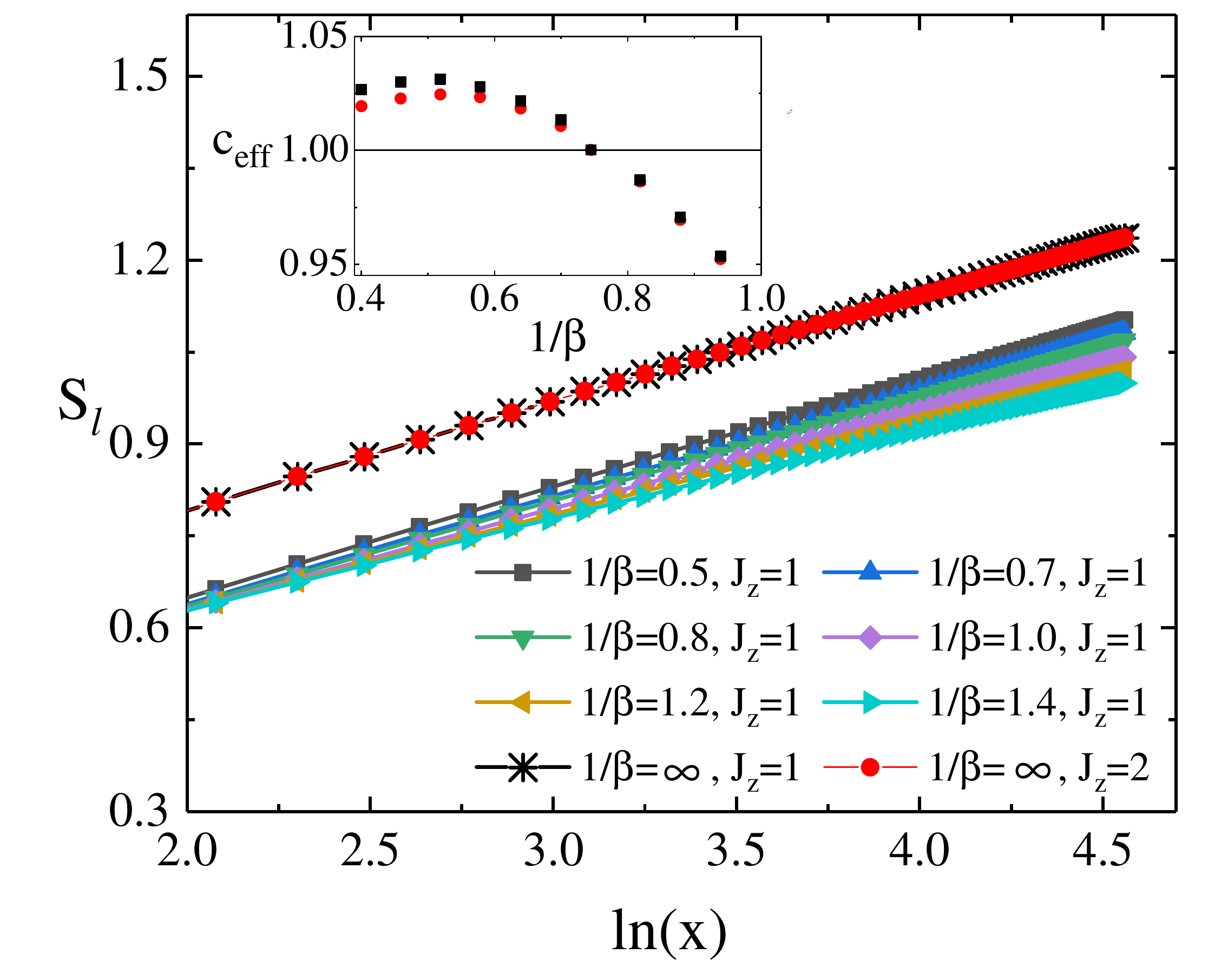}
\caption{\label{fig4} (Color online) The Scaling behavior of entanglement entropy versus  $\ln(x)=\ln[L/\pi\sin(\pi l/L)]$ for different values of $\beta^{-1}$ with $L= 300$. Inset shows the fitted coefficients as a function of $\beta^{-1}$ for system sizes $L=200$ (square) and $L=300$ (circle).
}
\end{figure}

Low-lying  excitation energy often reveals perspective features of different phases in the quantum many-body interacting systems. As mentioned previously, the system involves the gapless $XY$ phase for $-1<J_z\leq 1$ in the limit of $\beta=\infty$, which has the central charge $c_{\rm eff}=1$ owing to the conformal symmetry~\cite{Vidal03}. In the Jordan-Wigner representation of the Hamiltonian (\ref{Hamiltonian}), the spinless interacting fermion has a linear $1/L-$dependence for the {\color{red} finite-size} energy gap as a relativistic spectrum at the Fermi point or the low-lying property of the spectrum for the Luttinger liquid. When $\beta\neq \infty$, however, it is clearly of great interest whether such a $XY$ phase can be robust against a strong long-range repulsive interaction. For $J_z=1$ and $\beta=1$~\cite{Schulz, Li}, it was suggested that the ground state would be a quasi-Wigner crystal (WC), which results from the dominant long-range repulsive interaction over the kinetic energy. We calculated the {\color{red} finite-size} gap energy $\Delta(L)$ between the ground state and the first excitation energies as a function of system sizes for various cases as illustrated in Fig.\ref{fig3}, one can see that the energy gap $\Delta(\beta,J_z)$ can be either zero, including the case of  $J_z=1$ and $\beta=1$, or finite in the thermodynamic limit, which can be assigned to $XY$ and {\color{red} gapped quasi-}WC phases, respectively. However, for given $J_z$, when $\beta$ approaches its critical values $\beta_c$ from either $XY$ phase or WC phase where $\Delta(L)=\Delta(\beta,J_z)+A_1/L+O(1/L^2)$~\cite{You14}, it becomes rather difficult to accurately determine  the phase boundary between these two phases due to limited precisions on tiny values of $\Delta(L)$. Instead, we adopt the effective center charge $c_{\rm eff}$ deducted from the scaling behavior of the entanglement entropy given in Eq.(\ref{eqSl}) which enable us more accurately to allocate the phase boundary. We note that this scaling behavior is valid in the presence of the long range interaction as demonstrated numerically in Fig. \ref{fig4}, although it was originally derived for the short range interacting cases with conformal symmetries~\cite{Cardy, Laflorencie}.

Figure \ref{fig4} shows the entanglement entropy as a function of $\ln[L/\pi\sin(\pi l/L)]$ for various values of $\beta$ and positive $J_z$. It is instructive that the entanglement entropy still follows up the scaling behavior of Eq. (\ref{eqSl}), although conformal symmetries are not yet known here in general. Subsequently, the slope of the linear behavior gives rise to an effective central charge $c_{\rm eff}$ which  varies with $\beta$  as illustrated for system sizes $L=200$ and $300$ at $J_z=1$  in the inset of Fig. \ref{fig4}. One can see that finite-size effects for small $1/\beta$ is small but still visible, resulting in the correction to $c^0_{\rm eff}=1$ for the thermodynamic limit, but diminishes with increasing $1/\beta$. The curves for these two sizes cross with a horizontal line corresponding to $c_{eff}=c^0_{\rm eff}$ at $1/\beta_c=0.756$, where irrelevant corrections vanish to Eq. (\ref{eqSl}). The finite-size effect then becomes negligible for  $\beta\leq\beta_c $. This provides alternative way with higher accuracy to determine transition points between the $XY$ (critical) and WC (noncritical) phases~\cite{Alet,gong16,gong17}. 
\begin{figure}[h]
\includegraphics[width=0.48\textwidth]{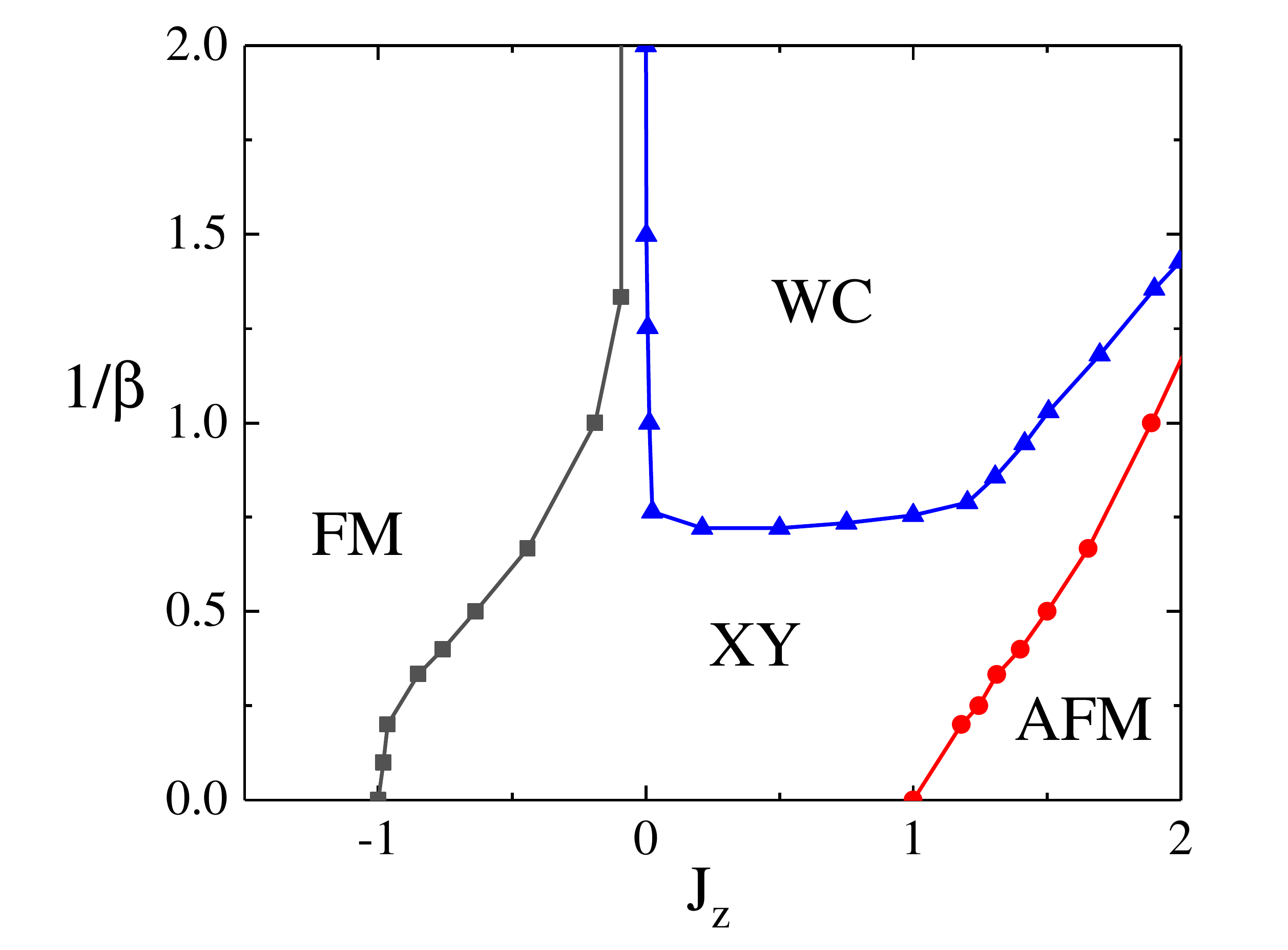}
\caption{\label{fig5} (Color online) Phase diagram of Hamiltonian (\ref{Hamiltonian}) as a functions of the interaction $J_z$ and  $1/\beta$ with $\alpha\rightarrow+\infty$. }
\end{figure}

In addition, we note that the FM phase is formed owing to the instability of effectively attractive density-density interaction for $J_z\leq 0$ upon changing $1/\beta$. Accordingly, the central charge is zero for the FM phase, but it has the value of 3/2 on its phase boundary with the $XY$ phase for the thermodynamic limits\cite{Chen,Olalla,Alba}.
To this end, the phase diagram is depicted in Fig. \ref{fig5} for $\alpha=\infty$. One can see that the critical points between the $XY$ phase and the FM phase asymptotically approach $J_z=0$, while the critical points between the AFM phase and the $XY$ phase mounts up with increasing $1/\beta$. Moreover, it is worthwhile to mention that at $\beta=0$ with $J_z>0$, $J_z$ term effectively results in one sort of long-range frustrations and has the same strength for all the sites, among which diagonal elements cancel each other in the ground state in correspondence to $S^z_{total}=0$ subspace\cite{Zerobeta}. In this case, the ground state again becomes gapless and the central charge equals to one. Particularly, the energy gap $\Delta(L)$ is scaled to zero in the limit of $L\rightarrow \infty$ independent of $J_z$ as illustrated for both $J_z=1$ and $J_z=2$ in Fig. \ref{fig3}. Moreover, the entanglement entropy behaves as same between $J_z=1, 2$, resulting in  $c_{\rm eff} \simeq 1.02$, as seen in Fig. \ref{fig4}. As connected to $J_z=0$, it is natural to consider that the system is indeed in the $XY$ phase, i.e. the transition between the FM and $XY$ phases takes place at $J_z=0$ for $1/\beta=\infty$.

\subsection{$1/\beta =0$}
In this section, we turn to the case of $\beta\rightarrow +\infty$. In this case, only the nearest neighbor interaction survives in the $J_z-$terms of the Hamiltonian Eq. (\ref{Hamiltonian}). The exponent $\alpha$ of the $XY-$long range interaction can be considered a tunable parameter to explore the quantum phase transition for various values of $J_z$.

\begin{figure}[h]
\includegraphics[width=0.48\textwidth]{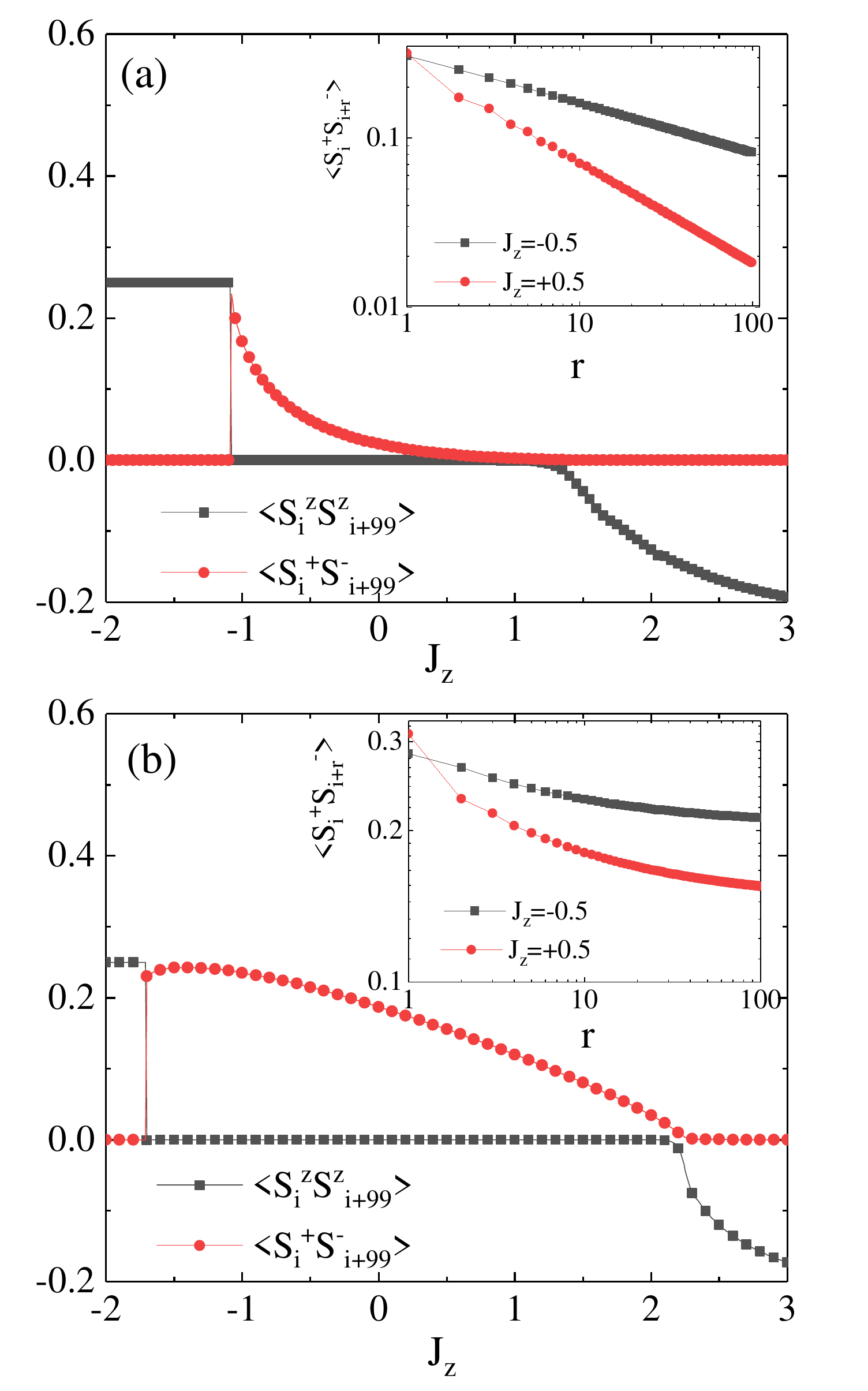}
\caption{\label{fig6} (Color online)  Correlation functions $\langle S^+_{i}S^-_{i+99}\rangle$ and $\langle S^z_{i}S^z_{i+99}\rangle$ are plotted as a function of the interaction $J_{z}$ for  (a) $\alpha=4$ and (b) $\alpha=2$. Inset: A log-log plot of  $\langle S^+_{i}S^-_{i+r}\rangle $ as a function of the distance $r$ with $J_z=\pm 0.5$.
 }
\end{figure}

Figure \ref{fig6} shows the dependence of two-spin correlations on $J_z$ with a distance of $|i-j|=99$
for different $\alpha$, calculated by using the iDMRG algorithm. When $J_z$ is negatively large enough,  $\langle S^+_{i}S^-_{i+99}\rangle=0$, $\langle S^z_{i}S^z_{i+99}\rangle =1/4$, suggesting that the system is in the FM phase.
When $J_z$ is sufficiently large, the transverse correlations remain zero, whereas $\langle S^z_{i}S^z_{i+99}\rangle$ becomes negative
so that the ground state is a AFM state. Analogous to the case of  $\alpha=\infty$, here we again utilize the discontinuity of the correlation functions to allocate the critical points for $\alpha$ and $J_z$ at the boundary of the FM phase, while the boundary of the AFM phase is also determined in terms of the entanglement entropy (see below).

In an intermediate range of $J_z$, one can further see that the transverse correlations $\langle S^+_{i}S^-_{i+99}\rangle$ is positive but longitudinal correlations $\langle S^z_{i}S^z_{i+99}\rangle$ vanish. Interestingly, we find that $\langle S^+_{i}S^-_{i+r}\rangle$ is a concave function of $J_z$ for $\alpha=2$, but becomes a convex one for $\alpha=4$.
Moreover, when $J_z=\pm0.5$, $\langle S^+_{i}S^-_{i+r}\rangle$ behaves as a power-law of $1/r$, vanishing in the limit of $r\rightarrow\infty$ as illustrated for $\alpha=4$ in the inset of Fig. \ref{fig6}(a), but ${\lim_{r \to +\infty}}\langle S^+_{i}S^-_{i+r}\rangle$ approaches a finite constant as seen for $\alpha=2$ from the inset of Fig. \ref{fig6}(b).
Therefore, the ground states for $\alpha=2$ in the intermediate range of $J_z$ is different that for $\alpha=4$.

In this range of $J_z$, it is natural to assign the large$-\alpha$ phase to the $XY$ phase, since this phase contains a special case where $\alpha=\infty$ and $J_z=0$ such that the Hamiltonian (\ref{Hamiltonian}) is reduced to describe a standard $XY$ chain, as already shown Fig. (\ref{fig5}). Moreover, when $\alpha$ is small or even not too large, one can show that a $U(1)$ symmetry in the ground state is spontaneously broken at $J_z = 0$ with using the conformal field analysis and perturbation calculation\cite{gong17}. It turns
out that one can expect the emergence of a continuous symmetry breaking (CSB) phase with gapless excitations for a small$-\alpha$ phase. It has been shown that a Berezinskii-Kosterlitz-Thouless like transition happens between the CSB phase and the $XY$ phase at $1/\alpha_c\simeq0.34$, at which the central charge is numerically increased by $4\%$ from unit. However, the criteria of the $4\%$ addition to the central charge might be invalid for the determination of the critical points with general values of $J_z$. To address this issue, we calculate the fidelity susceptibility which has been proposed for the identification of the critical points of continuous quantum phase transitions\cite{Gu2010} and even deconfined quantum critical points~\cite{Sun19}, and successfully applied to various strongly correlated systems~\cite{You15,You17,Ren18,Luo18}.

As a quantum information metric~\cite{Gu2010,You}, the fidelity measures the similarity between the two closest ground states when the parameter $\alpha$ is tuned tiny for the Hamiltonian (\ref{Hamiltonian}), which is defined as
\begin{equation}
\label{eq7}
F=|\langle\psi_0(\alpha)|\psi_0(\alpha+ \delta \alpha)\rangle|,
\end{equation}
where $\delta \alpha$ denotes a tiny deviation. Subsequently, we obtain the derivatives of interactions $\delta J_{i,j}=-\frac{J_{xy}}{|i-j|^\alpha}\ln |i-j| \delta \alpha$, where $J_{i,j}$ is the interaction strength between two spins at sites $i$ and $j$. The average derivatives of interactions per site are practically considered as an effective tuning parameter $\delta J=\frac{\sum_{i<j}\delta J_{i,j}}{L}$. Therefore, the fidelity susceptibility per site can be calculated numerically by
\begin{equation}
\label{eq5}
\chi=\lim_{\delta J\rightarrow
0}\frac{-2 \textrm{ln}F}{L(\delta J)^2},
\end{equation}
whose peak is thus used to identify the critical value of $\alpha$ and to separate the CSB phase from the $XY$ phase for each $J_z$.

\begin{figure}[h]
\includegraphics[width=0.51\textwidth]{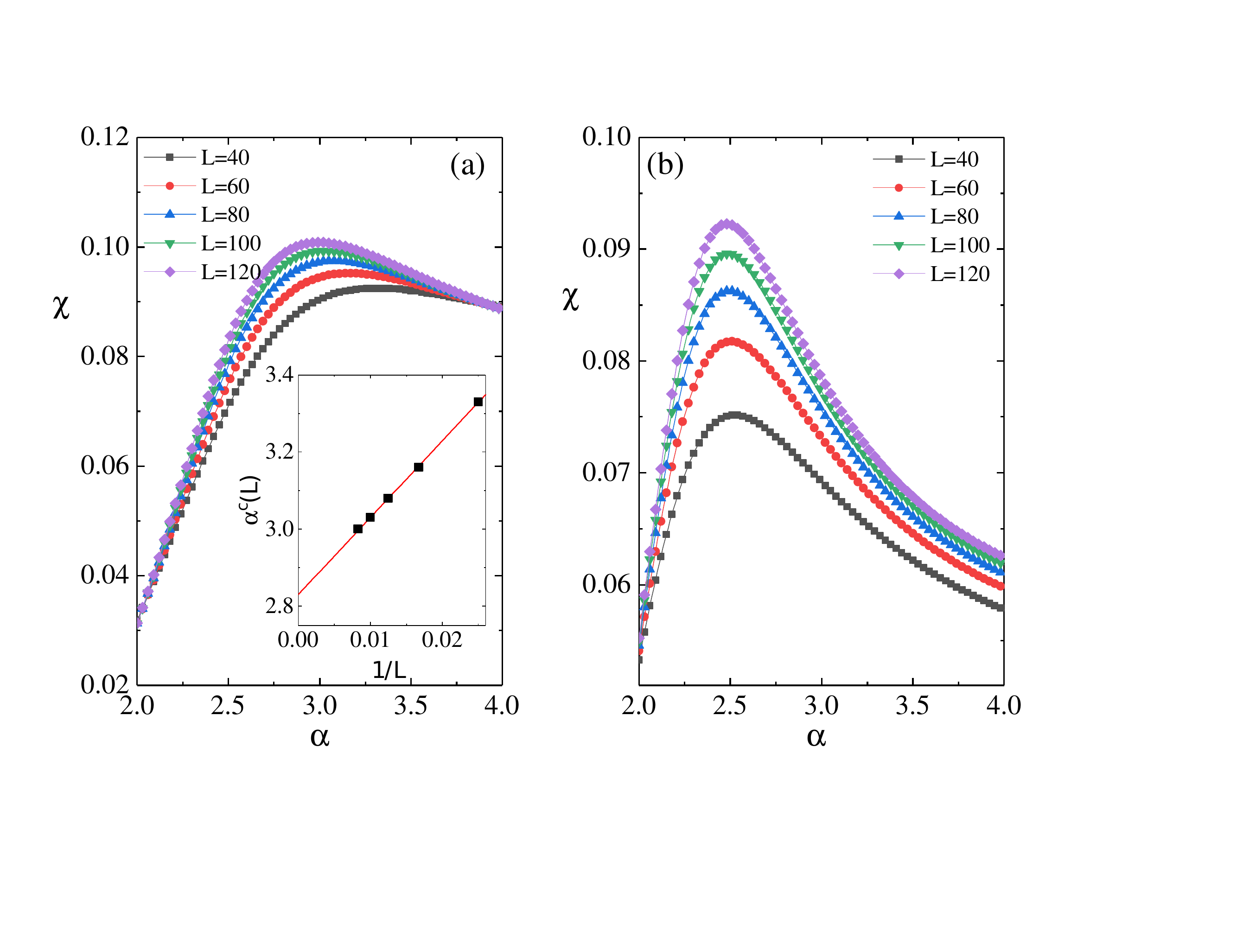}
\caption{\label{fig7} (Color online) Fidelity susceptibility per site is plotted as a function of parameter $\alpha$ for various system sizes with (a) $J_z=0$ and (b) $J_z=1.0$. Inset: Scaling behavior of the fidelity susceptibility peak points with respect to $1/L$. }
\end{figure}

In our numerical calculations, we take $\delta \alpha=0.005$. For the case of $L=100$ and $\alpha=3$, the effective tuning parameter $\delta J\simeq 0.001$.  The ground-state fidelity susceptibility per site $\chi$ is shown for $J_z=0,1$ as a function of the parameter $\alpha$ for different sizes in Fig. \ref{fig7} (a) and (b), respectively. For each $J_z$, one can see that the peaks of $\chi$  grow with respect to increasing the system size  so that a divergence peak would be expected for the $L\rightarrow\infty$ limit to signal the appearance of a quantum phase transition. In order to locate the quantum critical point $\alpha_c$ for the thermodynamic limit, we uses the finite-size scaling analysis to obtain $\alpha_c=2.83$ and $\nu=1$ at $J_z=0$ as seen in the inset of Fig. \ref{fig7}(a). This value of $\alpha_c$ is good consistent with that determined by the central charge and the perturbation theory calculation~\cite{gong17}. Similarly, we can determine critical points at other values of $J_z$ for the boundary between the CSB and $XY$ pases.  In particular, the critical value of $\alpha_c=2.45$ for $J_z=1.0$ is obtained from the results shown in Fig. \ref{fig7}(b).

\begin{figure}[h]
\includegraphics[width=0.51\textwidth]{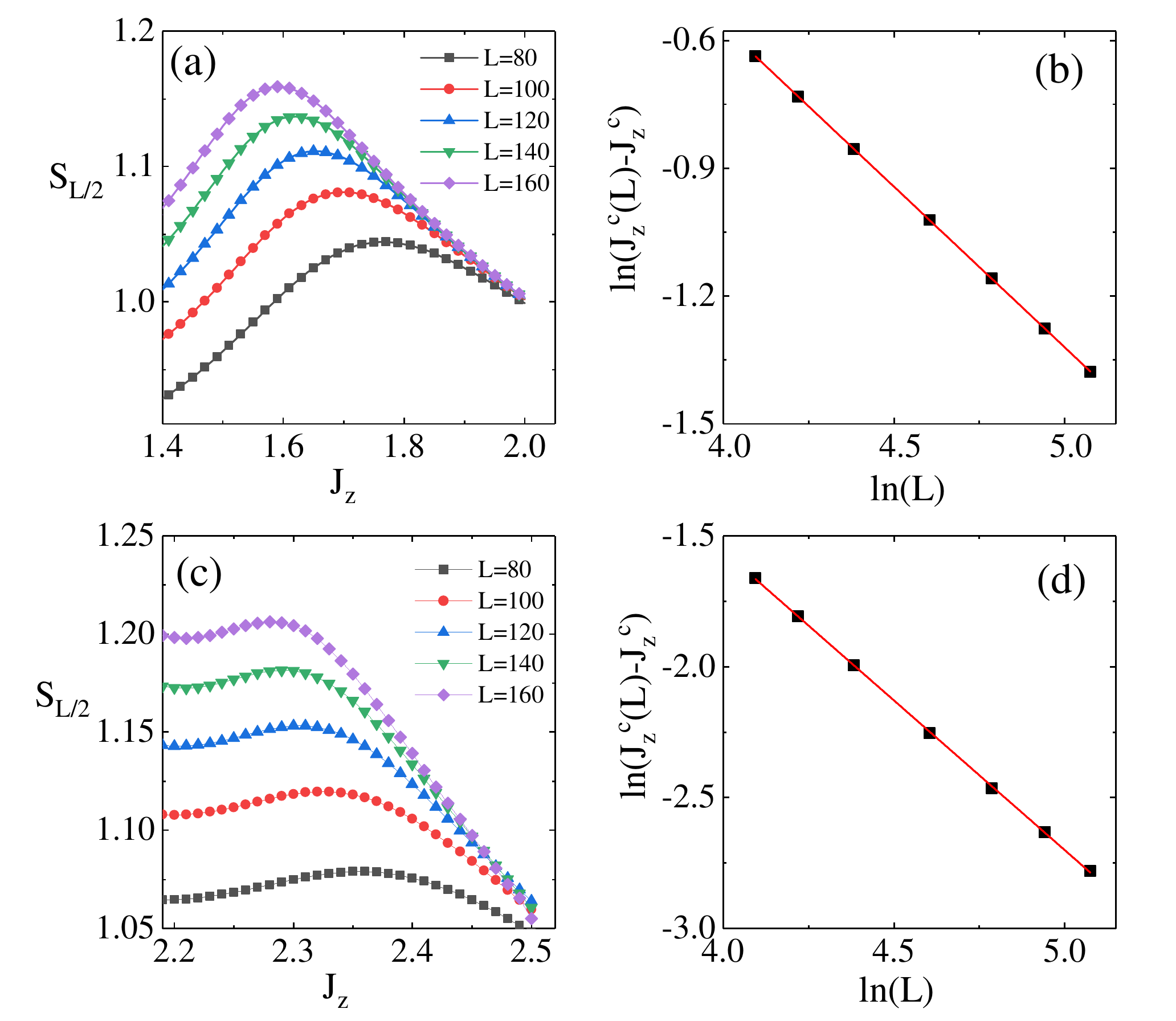}
\caption{\label{fig8} (Color online)  Entanglement entropy is plotted as a function of the interaction $J_{z}$ on different system sizes for (a) $\alpha=4$ and (c) $\alpha=2$.  The peak positions of $S_{L/2}$ versus the system size $L$ for (b) $\alpha=4$ and (d) $\alpha=2$. }
\end{figure}

Now we turn to quantum phase transitions between the intermediate and AFM phases, which are characterized by the peaks of the entanglement entropies as demonstrated for $\alpha=2,4$ in Fig. \ref{fig8}. One can see that the peaks for both cases in (a) and (c) of Fig. \ref{fig8} move to lower values of $J_z$ when $L$ increases. Fitting the locations of peaks with the formula (\ref{eq3}) as shown in (b) and (d) of Fig. \ref{fig8}, one can obtain that $J_z^c=1.35$ and $2.21$, respectively. Such fitted results agree very well with the inflexion points of the correlations shown in Fig. \ref{fig6}. In the same manner, we allocate more critical values of $J_z$ and $\alpha$ for the boundary of the AFM phase with both $XY$ and CSB phases.

\begin{figure}[h]
\includegraphics[width=0.45\textwidth]{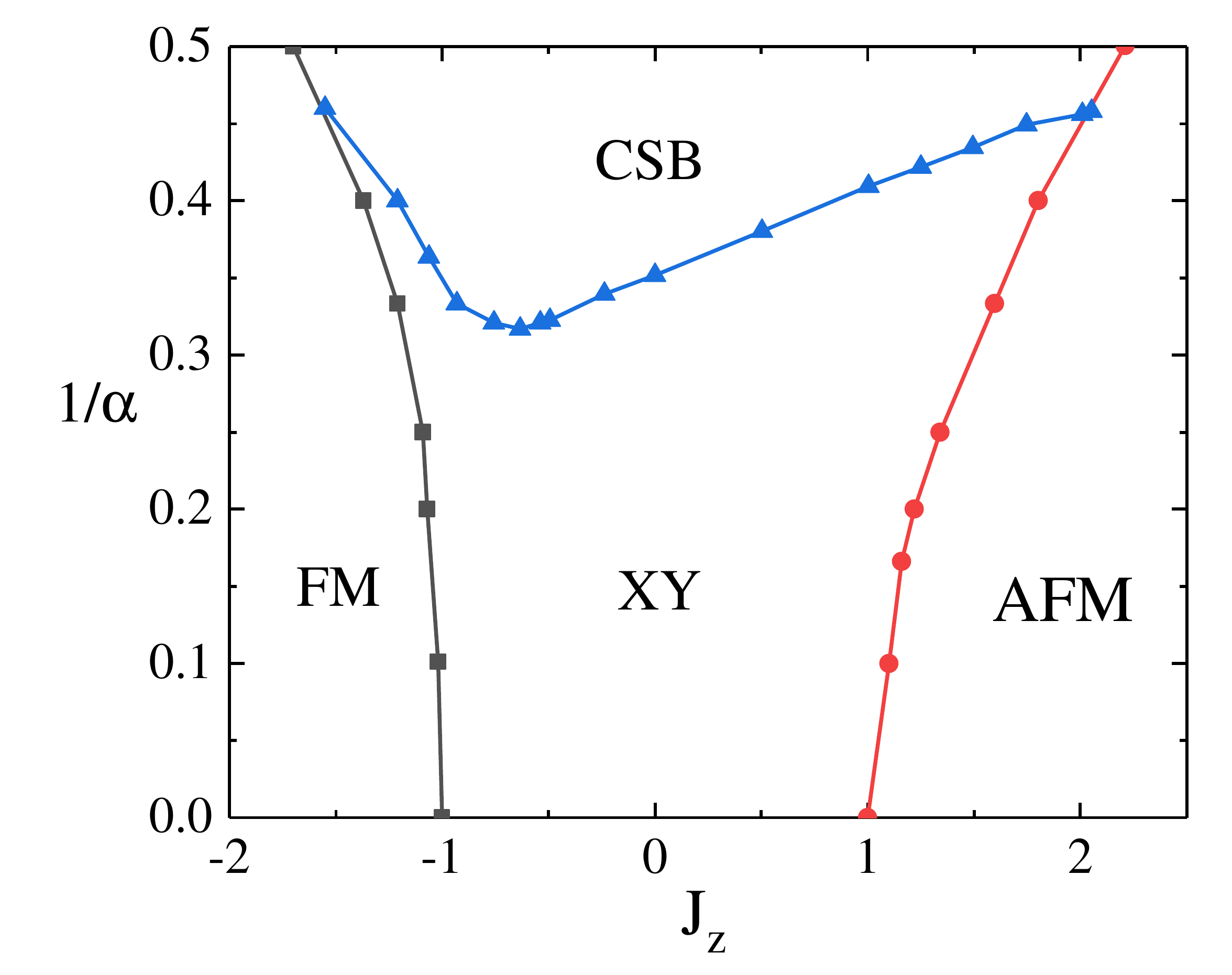}
\caption{\label{fig9} (Color online)  Phase diagram of Hamiltonian (\ref{Hamiltonian}) as a functions of the interaction $J_z$ and $\alpha$ with $\beta\rightarrow+\infty$. }
\end{figure}

Based on the above analysis on the properties of the correlation functions, the fidelity susceptibility and the entanglement entropy, we establish the ground-state phase diagram for the Hamiltonian (\ref{Hamiltonian}) with $\alpha=\infty$ as shown in Fig. \ref{fig9}.

\vspace{0.3cm}
\section{Discussion}
\label{sec:Discussion}

In this paper, we study quantum phase transitions for a quantum spin-$1/2$ chain with anisotropic power-law-decaying long-range interactions, which are characterized by exponent parameters $\alpha$ for $xy-$term and $\beta$ for $z-$term, by employing density-matrix renormalization-group method. With numerically analyzing the effects of $\alpha$ and $\beta$ on the spin-spin correlation functions, the entanglement entropy and the central charge, and the fidelity susceptibility, we establish two phase diagrams for $\alpha=\infty$ and $\beta=\infty$, respectively.

Both cases involve a ferromagnetic phase and an antiferromagnetic phase corresponding to sufficiently negative and positive $J_z$, respectively. However, in the intermediate regime of $J_z$, the former involves not only a usual $XY$ phase effectively equivalent to a short range repulsive density-density interaction, but also a Wigner-crystal phase which essentially results from for a sufficient strong long-range $J_z$ term; for the later, the {\color{red} gapped} Wigner crystal phase is replaced by a continuous $U(1)$ symmetry breaking phase. Moreover, it is interesting to notice that the WC and CSB phases actually reveal two different mechanisms, which intrinsically result from either two-body processes of the strong long-range repulsive interaction or one-body kinetic processes of the long-range hoping in the fermion representation.

From this study, we found that the entanglement entropy and the central charge can be used efficiently to extract critical values of the quantum phase transition between two phases when one of them possesses a well-defined central charge but another one is gapful \cite{Luo2019}. However, when one is encountered with a quantum phase transition between two gapless phases, the fidelity susceptibility alternatively provides a more feasible way to allocate the critical points as applied to the transition between the $XY$ and continuous $U(1)$ symmetry breaking phases.

We so far focus on the ground state phase diagrams only for $\alpha=\infty$ and $\beta=\infty$. There are actually a couple of important aspects beyond the above two cases for the Hamiltonian (\ref{Hamiltonian}), such as ground phase diagrams with $\alpha=\beta$ and $J_{xy}>0$, extensions to two leg-ladders and even two dimensions, etc. The emergence of any non-trivial gapless phase, corresponding novel low-lying excitation spectra or exotic collective excitations with special symmetries,  and thermodynamic and dynamic properties would be very interesting questions for the presence of long range interactions but are certainly open for further studies in the future.

\begin{acknowledgments}
This work is supported by the National Program on Key Research Project (Grant No. 2016YFA0300501) and the National Natural Science
Foundation of China under Grants No. 11104021, 11474211, 61674110 and 11974244. W.L.Y is appreciative of support from the start-up fund of Nanjing University of Aeronautics and Astronautics. X.W. also acknowledges additional supports from a Shanghai talent program.
\end{acknowledgments}


\begin{thebibliography}{25}
\bibitem{Zerobeta} When $\beta=0$, $J_z$ term in Eq. (\ref{Hamiltonian}) can be written as $J_z[(S^z_{total})^2-L/4]$ under the periodic boundary condition. For small values of $\beta$, whether effective cancelling of those diagonal elements remains is to be further explored.
\end{thebibliography}
\end{document}